# *Proton Flows, Proton Gradients and Subcellular Architecture in Biological Energy Conversion*

*Luca Quaroni*

*Faculty of Chemistry, Jagiellonian University, ul. Gronostajowa 2, 30-387, Cracow, Poland*

luca.quaroni@uj.edu.pl

ORCID: 0000-0002-0225-9775

**Abstract**

Hydrogen cations, or protons, provide the medium by which energy is stored and converted in biological systems. Such pre-eminence relies on the interplay between interfacial and bulk chemical transformations, according to mechanisms that are shared by organisms in all phyla of life. The present work provides an introduction to the fundamental aspects of biological energy management by focusing on the relationship between vectorial proton flows and the geometry of energy producing organelles in eukaryotes. The two leading models of proton-mediated energy conversion, the delocalised proton (or chemiosmotic) model and the localised proton model, are presented in a complementary perspective. While the delocalised model provides a description that relies on equilibrium thermodynamics, the localised model addresses dynamic processes that are better described using out-of-equilibrium thermodynamics. The work reviews the salient aspects of such mechanisms, traces the development of our present understanding, and highlights the areas that are open to future developments.

**Keywords**

Proton flows, Chemiosmosis, Localized Protons, Delocalized Protons, Energized Membranes

**Competing Interests**

The author declares no competing interests.

**Acknowledgements**

The author thanks his wife for the support provided during the writing up of the manuscript.

## *Main Text*

### *1.0 Vectorial Processes in Cellular Biology*

Our familiarity with most biochemical processes comes from reactions that are reproduced in the laboratory as bulk experiments, with reactants homogeneously dissolved in solution. However, our experience is biased by the easy accessibility of such *in vitro* demonstrations, and this is not necessarily how reactions occur *in vivo*. Inside living cells, membranes, organelles and interfaces~~,~~ create a spatial architecture where the concentration of molecules and ions can evolve in space and time non homogeneously. The effect of compartmentalization goes beyond that of confining reactions to specific locations. In such a fragmented environment, many processes are *vectorial,* whereby the consumption of chemical species in one location is accompanied by the formation of new species in a separate location. Reaction turnover is coupled to displacement. Extending the mathematical analogy, homogenous bulk reactions can instead be described as *scalar*. (Figure 1) Vectorial transformations involve the formation, remodelling and depletion of concentration gradients, which are accompanied by corresponding gradients in the electrochemical potential of reactants and products. (Figure 1) Energy is absorbed to drive molecules and charges against their concentration gradient and energy is released when they move back along the concentration gradient. Because of the non-uniform distribution of electrochemical potentials, vectorial transformations provide a connection between chemical composition, energy and space, and are at the core of biological energy management. (Mitchell, 1962)

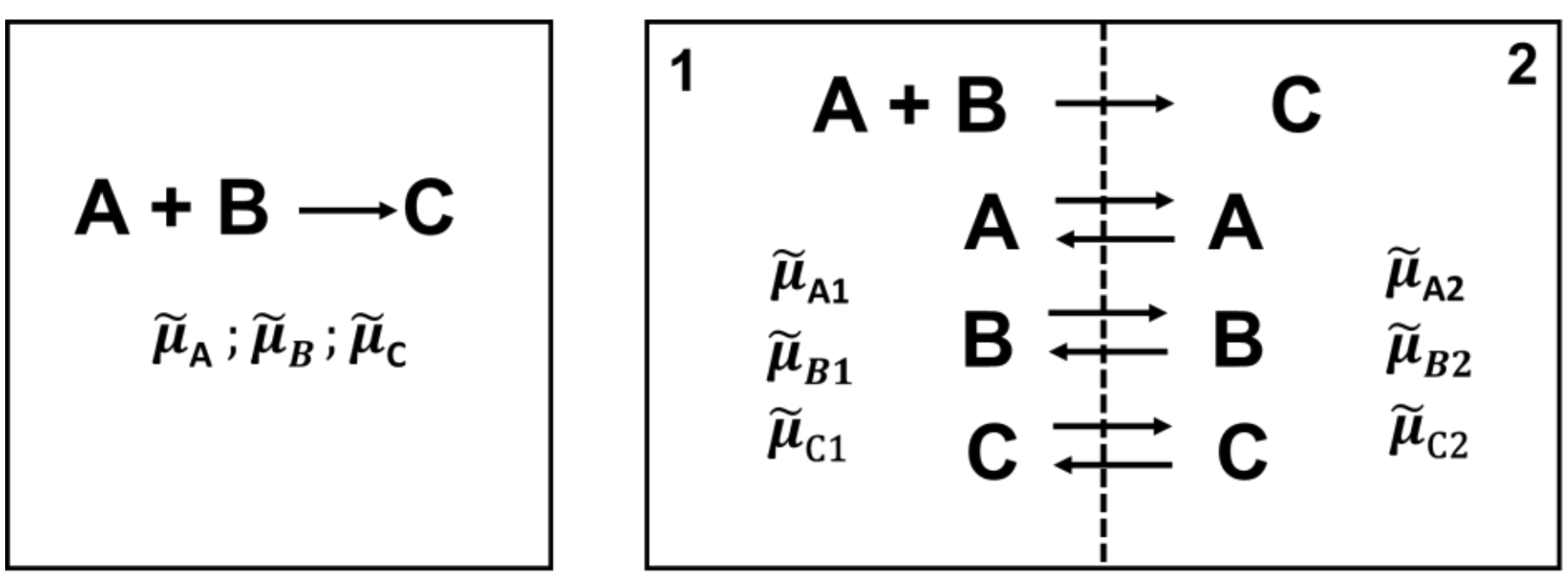

***Fig. 1 Scalar Reactions vs. Vectorial Reactions.*** *Left. In a scalar reaction, reactants (A, B) and products (C) are uniformly distributed within the same volume of space. Each species is associated to one value of the electrochemical potential. Right. In a vectorial reaction reactants are consumed in one region of space and products are formed in a different region of space. Different values of the electrochemical potential for each compound are associated to different regions. Gradients in electrochemical potential can drive diffusion processes or can store energy as a potential difference*

Vectorial reactions are enabled by cellular membranes, which define organelles, enclose compartments and the cytoplasm itself, and modulate flows of molecules, electrons and ions across them and along their surface. Vectorial processes that generate charge separation are particularly relevant, because the contribution from electrostatic energy is markedly sensitive to small changes in spatial distribution. Many biological entities are charged, from proteins and nucleic acids to small molecules. However, the ones that dominate cellular energy interconversion are the two smallest ones, the electron and the hydrogen cation, $H^+$, i.e. the proton.

Historically, the role of vectorial electron transfer in cellular energy management was recognized early. In contrast, the role of vectorial proton transfer and the formation of proton gradients initially received less attention. Recognition came in the second half of the $20^{th}$ century, following investigations on the sequential transformations that lead to the synthesis of ATP in oxidative and photosynthetic phosphorylation, which also revealed the interplay of electron and proton transfer in cellular biochemistry. In oxidative phosphorylation, electrons are extracted from low redox potential molecules and transferred to dioxygen via a chain of redox reactions taking place along the inner mitochondrial membrane. The energy is eventually used to form ATP, which is later hydrolysed to drive a multitude of cellular processes. In photosynthetic phosphorylation, it is the energy from light absorption that triggers the chain of reactions leading to ATP synthesis.

Both processes are strictly associated to the membranes of specific organelles, and ATP synthesis would not occur in bulk suspension. However, the underlying reason was still unknown at the end of the 1950's. A widespread expectation at the time was that a short lived molecular species, a "high energy intermediate", was created in the early stages of oxidative or photosynthetic phosphorylation, and would then undergo rapid conversion to the ATP molecule, which acts as a more stable energy reservoir for cellular metabolism. (Mitchell, 2011) In such a scenario, the membrane has no apparent function, and the failure to isolate or even observe such molecular intermediate stimulated a search for alternative models.

A new approach was suggested by the analogy with vectorial charge transfer in fuel cells, which was already well understood. (Mitchell, 1979) Similarly to fuel cells, electron transfer in oxidative phosphorylation had to be associated to proton transfer between the aqueous phases adjoining a membrane. The role of protons provided the missing connection between dioxygen reduction and ATP synthesis. Two proposals, by P. Mitchell and R.J.P. Williams, emerged simultaneously in the early 1960's suggesting that the energy obtained by oxygen reduction could be transmitted by coupling vectorial electron and proton transfer at a membrane or at an interface (Williams initially used the word *displaced*, instead of *vectorial*). (Mitchell, 1961) (Williams, 1961) Both of them bypassed the formation of a high-energy molecular intermediate by attributing its role to local differences in proton concentration. Both of them encountered initial resistance and generated intense discussions. Acceptance grew only gradually, over the years, with the publication of corroborative experiments, and with the general progress in the understanding of membrane physiology.

### *1.1 The chemiosmotic model (or the delocalised proton model): energy conversion at thermodynamic equilibrium.*

In the early proposal by Mitchell, the role of vectorially transferred protons is to contribute to the equilibrium of ATP formation and hydrolysis. (Mitchell, 1961) Equation 1 shows one of the possible stoichiometries of the reaction, where $P_i$ represents the phosphate anion in accessible ionization states. The ionization state of ADP, ATP and $P_i$ depends on pH, with at least two ionization states accessible for each ion at physiological pH. ATP synthesis and hydrolysis are catalysed by ATP-synthase, an integral membrane protein. Mitchell postulated an asymmetric access to the active site of ATP-synthase for protons and from the adjoining aqueous phases. The direct consequence of this topology is that differences in acidity between the two phases can control the thermodynamics of the reaction, and drive it towards ATP synthesis or ATP hydrolysis. Therefore, the indispensable requirement of a membrane for ATP synthesis is explained by its role as a diffusion barrier, whereas the role of the early stages of oxidative phosphorylation is to generate the transmembrane difference in proton concentration by a series of vectorial proton-coupled electron transfer reactions.

$$ADP_i + P_i + 2H^+ \rightleftarrows ATP_i + H_2O \qquad \text{Equation 1}$$

The initial formulation of the model evolved over time. The observation that the ratio between consumed protons and synthesised ATP is unrelated to the stoichiometry of Equation 1 indicated that the role of the proton is not simply that of a reactant. Elucidation of the mechanism and structure of ATP-synthase revealed that the proton gradient is instead used to drive the conformational changes of the protein that result in ATP synthesis. (Kagawa, 2010)

In its mature formulation, the model became a general paradigm for cellular energy management that extends beyond oxidative phosphorylation. Its fundamental tenet is that the difference in proton concentration between the two aqueous phases adjoining a membrane, i.e. a proton gradient, is a medium for the storage of energy, which can be later used to drive cellular biochemistry.

The energy for creating the gradient can originate from exergonic chemical transformations, such as oxygen reduction in mitochondrial terminal oxidation, or from photon absorption, such as in the photosynthetic apparatus of the chloroplast or in the purple membrane of *Halobacterium salinarum*. Alternatively, protons can be consumed on one side of the membrane by a scalar process and released on the opposite side by a second, coupled, scalar process, resulting on balance in a vectorial translocation, as exemplified by quinone/hydroquinone loops. (Figure 2) The stored energy can later be used to drive a range of endergonic cellular processes, in addition to ATP synthesis, such as the vectorial translocation of other ions (both anions and cations) and molecules across membranes against their concentration gradient. (Mitchell, 2011) ATP itself can be used by an ATP-synthase operating in reverse, as an ATP-hydrolase (ATP-ase), to promote an endergonic transformation or to generate a gradient of protons in cellular locations removed from the original source. Energy can also be consumed unproductively, by uncoupling proton source and sink while allowing the proton gradient to dissipate, a process contributing to thermogenesis. (Nicholls 2023) The side of the membrane to which protons are actively translocated is termed the p-side, whereas the one from which protons are extracted is termed the n-side. In recognition of their role in energy conversion, protons accumulated on the p-side are sometimes called energized protons, while the membranes that support proton gradients are variously termed energized membranes, energy-transducing membranes, or energy-conserving membranes. Proteins that increase the proton concentration on the p-side are termed proton pumps or proton sources. The ones that use the energy stored by the protons are termed proton consumers or proton sinks. (Figure 2)

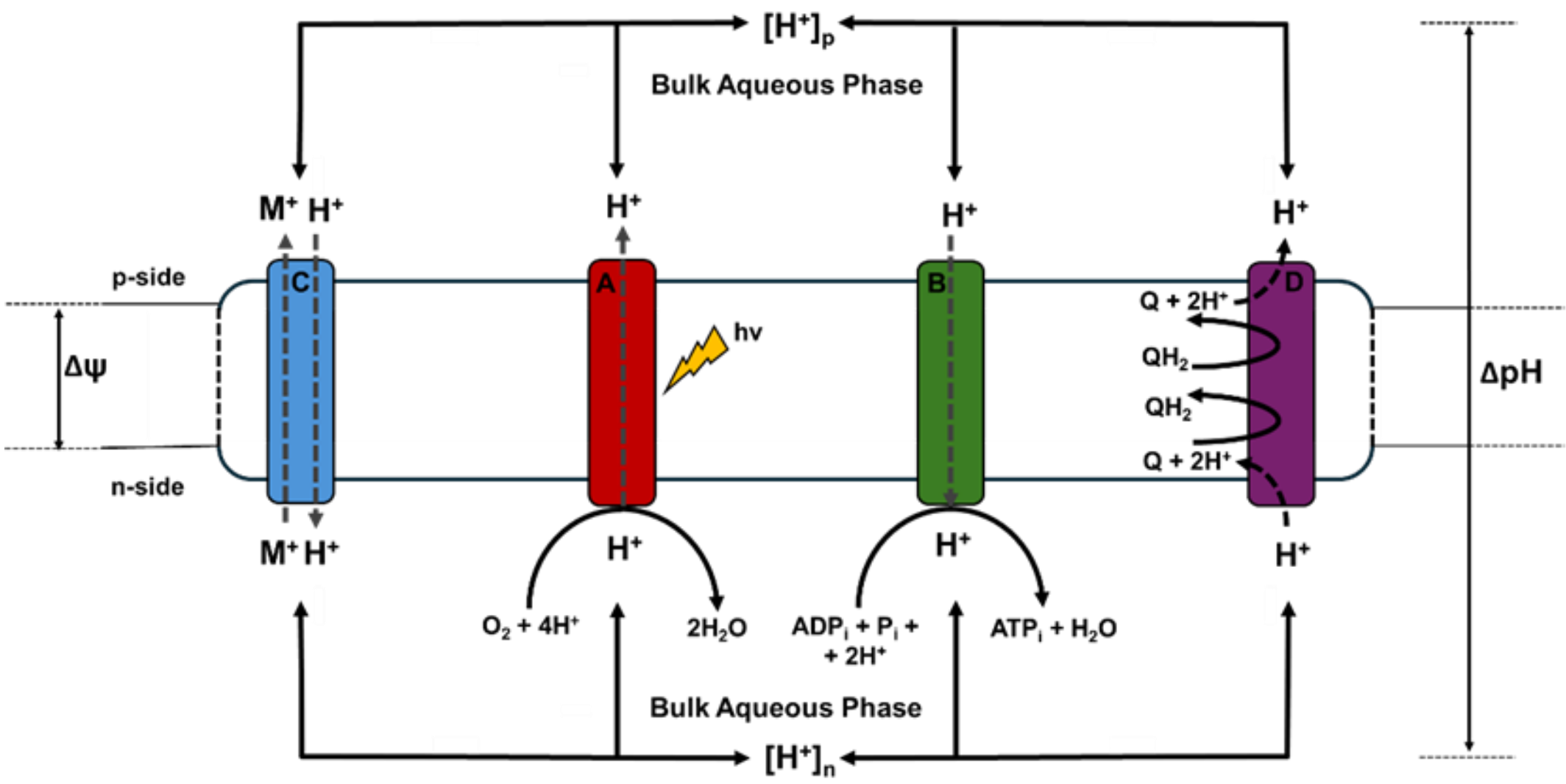

***Fig. 2 Proton gradient creation and consumption according to the chemiosmotic (delocalised proton) model.*** *The proton concentration is decreased at the n-side and increased at the p-side by chemical reactions (e.g. oxygen reduction) and proton translocation at proton pump A. Proton translocation can be driven by an exergonic chemical reaction or a physical process (light absorption). Translocation can also occur because of coupled proton/electron transfer reactions (proton pump D). Protons transferred to the p-side equilibrate with the bulk aqueous phase, contributing to the electrochemical potential. Some protons remain associated with the membrane as unpaired charges and contribute to the membrane potential. A proton gradient between the two opposite aqueous phases is used to store energy, which is extracted by returning protons from the p-side back to the n-side, thus driving ATP synthesis by ATP-synthase (proton consumer B) or the translocation of another cation by an antiport (proton consumer C)*

The model is quantitatively represented by equilibrium thermodynamics using the mass and energy balance between two aqueous phases separated by a membrane. The energy accumulated by the formation of proton gradients and associated ionic gradients is expressed by the electrochemical potential of the proton, $\tilde{\mu}_{H^+}$, and by the electrostatic transmembrane potential, $\Delta\psi$. $\tilde{\mu}_{H^+}$ takes on the role of high energy intermediate, in place of a molecular species. Differences in proton concentration between the two sides of the membrane result in a protonmotive force, *pmf*, in units of Volts, with an associated difference in electrochemical potential, $\Delta\tilde{\mu}_{H^+}$ (Equation 2). (Mitchell, 2011) Equation 2 is based on the assumption of a Donnan equilibrium for protons (Donnan 1924), whereas the transmembrane distribution of charged species contributes to the membrane potential $\Delta\psi$.

$$pmf = \frac{\Delta\tilde{\mu}_{H^+}}{F} = \Delta\psi + (2.3RT/F)\log(a_{H+p}/a_{H+n}) \qquad \text{(Equation 2)}$$

*R* is the gas constant; *T* is the absolute temperature; *F* is Faraday's constant; $a_{H+p}/a_{H+n}$ is the ratio of the activity of the proton between the two sides of the membrane, with $\log(a_{H+p}/a_{H+n}) = -\Delta pH$, the difference in pH between the two phases. Equation 2 is also applicable to dynamic processes (e.g. during proton consumption by ATP-synthase) provided that changes are slow enough to allow for the rapid re-establishment of equilibrium. Expected values of *pmf* during metabolic turnover are typically between 150 mV and 250 mV. (Mitchell 1969a; Kühlbrandt 2019) The *pmf* can be decreased by the presence of anions or other species that form neutral ion pairs or molecules capable of passive diffusion through the membrane. These species act as uncouplers of phosphorylation from the electron transfer chain and can slow or arrest phosphorylation by draining the proton gradient. (Mitchell 1969b)

Most protons introduced into one of the aqueous phases are balanced by counterions, and contribute to $a_{H+}$ while concurrently maintaining an electroneutral bulk phase. In contrast, local charge separation is observed at the membrane surface, in the region of the electrical double layer, where unpaired protons together with other ions and fixed membrane charges contribute to $\Delta\psi$. (Koch 1986) The more abundant cations in biological fluids, such as $K^+$, $Na^+$, $Ca^{2+}$, $Mg^{2+}$, can contribute to $\Delta\psi$, depending on the specific physiological context. $K^+$ is generally considered to be a major contributor to mitochondrial *pmf* because of its intracellular abundance, as supported by the early experiments by Mitchell and Moyle. (Mitchell 1969a) The exchange between $H^+$ ions and cations in the double layer region accounts for the functional equivalence of $\Delta\psi$ and $\Delta pH$, which can both enable ATP synthesis. In chloroplasts, the measured contribution from $\Delta\psi$ is minor, whereas large pH differences account for most of the *pmf*. In mitochondria the situation is reversed, and $\Delta\psi$ is the dominant contribution. (Kühlbrandt 2019)

Because of the analogy with the thermodynamic description of osmotic pressure, the model has been named the *chemiosmotic model*, and the associated processes are collectively called *chemiosmosis*. The model was later also termed the *delocalised proton model*, because proton diffusion from the membrane to the bulk and from the bulk to the membrane are necessary steps.

Because of the quantitative relationship provided by Equation 2, the chemiosmotic model was amenable to direct validation from its inception. Early experiments supported its validity by showing that the *pmf* at the mitochondrial inner membrane is correlated to metabolic activity, such as oxygen consumption and ATP synthesis. (e.g. Reid, 1966) Considerable momentum was provided by the independent "acid bath" experiments performed by Jagendorf on chloroplasts (Jagendorf, 1966), showing that a rapid increase of the external pH of thylakoids triggered the synthesis of ATP from ADP and phosphate. Analogous supporting experiments were later reported by multiple groups that achieved ATP synthesis or hydrolysis by controlling the pH gradient across ATP-synthases reconstituted in vesicles and planar bilayers. (Kagawa, 2010) As an increasing number of experimental results became available, the following decade saw progressive acceptance of the model.

### *1.2 The delocalised proton model: energy conversion away from equilibrium.*

In a separate proposal, Williams provided a model suggesting that the core of cellular energy management is the direct movement of protons from a proton source, typically a protein, to another protein that acts as the proton sink along a membrane. (Williams, 1961) In agreement with the chemiosmotic model, it recognizes the role of vectorial processes, including proton transfer, in cellular energy management. In contrast to the chemiosmotic model, it provides the conceptual framework to interpret dynamic processes. In earlier formulations, it focused on charge transfer along chains of membrane proteins, whereby proton movement occurs via a shuttling mechanism inside the proteins and the membrane itself (Figure 3A). Later formulations considered flows of protons localized at the membrane-water interface, together with the better understood transmembrane flows (Williams, 1978a; Williams, 1978b). The model has been termed the *localised proton model*, because protons remain associated to a membrane or interface throughout their transfer from source to sink, without the need to equilibrate with bulk phases. While the localised model bypasses the requirement for a high energy molecular intermediate, it also refrains from involving the electrochemical potential of the proton as a physical counterpart to it. Its core element is the connectivity of source and sink, enacted either by proximity of the two or by their enhanced mobility within the membrane. In either case, both of them are located at the same interface, a requirement which is absent from the chemiosmotic model.

The localised proton model has been presented as a more general description of the proton circuits that mediate cellular energy interconversion. The three phase architecture that is the basis of the delocalised model is not a requirement for creating proton gradients. It is the direct transfer of protons between source and sink that is both necessary and sufficient for the purpose of energy transmission. Accordingly, an interface between two phases is the only required structure, providing a surface that connects source and sink via proton diffusion (Figure 3B). (Williams, 1978a; Williams, 1978b)

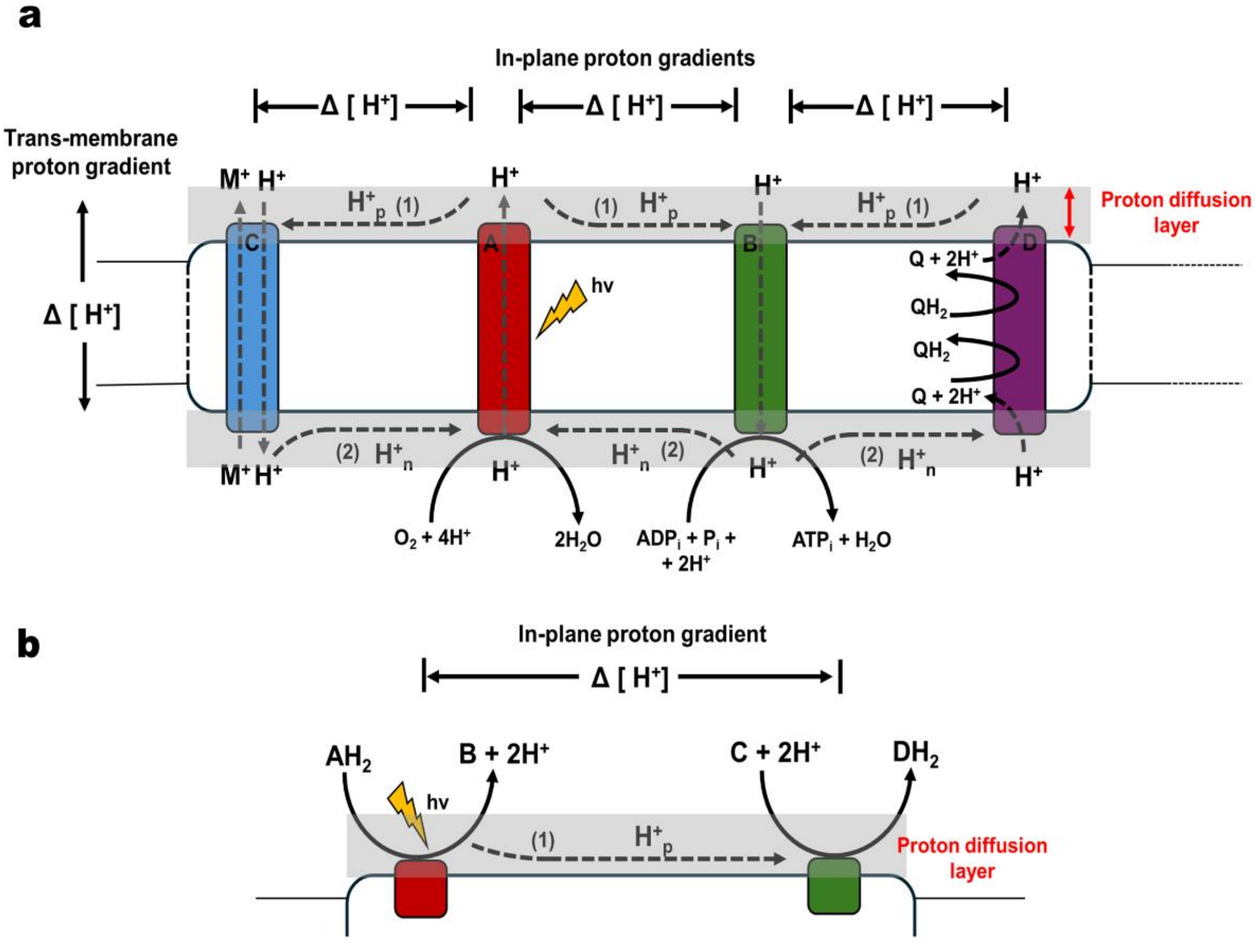

***Fig. 3 Proton gradient creation in the membrane plane according to the localized model.*** *a. The same processes as in Figure 2 translocate protons from the n-side to the p-side. However, protons do not equilibrate with bulk aqueous phases. but remain associated to the membrane in their movement between proton sources and proton consumers. Proton gradients are formed both across the membrane and between sources and consumers, without diffusion to the bulk aqueous phase. b. Minimal case of localized proton gradient creation. Protons accumulate at one location of an interface, not necessarily a membrane, by the (physical or chemical) action of a proton source and are consumed at a different location by the action of a consumer. A proton gradient is created in the proximity of the surface between the two. Transport across the substrate and diffusion to the bulk are unnecessary*

The initial lack of a direct quantitative representation of the localised model limited opportunities for early validation. Nonetheless, the model is consistent with experimental observations that are discordant with the chemiosmotic model and cannot be accounted for by Equation 2. (Ferguson 1985) Microbial systems provided an early point of contention. It was argued that chemiosmosis could not explain microbial energetics because protons released to the large external aqueous phase would provide a negligible contribution to $\tilde{\mu}_{H^+}$ and require extremely long diffusion times, akin to dilution into the Pacific Ocean. (Williams, 1978a) The argument was affected by the limited understanding of the bacterial envelope in the early 1960's. Nonetheless, cases were reported where ATP synthesis could take place despite insufficient *pmf* and unfavourable *ΔpH*. Cells of the archaeon *Halobacterium salinarum* were shown to synthesise ATP with pH as high as 8 on the p-side. (Michel 1980) Similarly, some gram-positive alkaliphilic bacteria can produce ATP in an environment with even higher external pH, up to pH ~10, despite adverse values of $\Delta pH$ and insufficient *pmf*. (Krulwich, 1995; Hicks, 1995) Localised proton transfer can justify these observations by bypassing the need for equilibration to generate *pmf*.

### *1.21 Out-of-Equilibrium Thermodynamics of Delocalised and Localised Proton Transfer*

Quantitative analysis of localised proton transfer has been allowed by the application of out-of-equilibrium thermodynamics, replacing the equilibrium description provided by Equation 2. Fluxes of energy and matter are described as functions of gradients or differences in thermodynamic variables. In this respect, they provide a tool that is better suited than equilibrium thermodynamics to the study of living systems, which are inherently out-of-equilibrium. The formalism is particularly amenable to describe vectorial processes, providing quantitative relationships between the force that drives a transformation and the magnitude of the resulting changes (flow-force relationships). (Westerhoff, 1988) The simple one-dimensional case of flow-force relationship is shown in Equation 3, where $J$ represents a flux (e.g. particles per unit area per unit time), $\Delta q$ is a gradient that acts as a driving force, and $L$ is a phenomenological kinetic coefficient.

$$J = L\,\Delta q \qquad \text{Equation 3}$$

When applied to cellular energy conversion, $\Delta q$ can correspond to $\Delta\psi$, $\Delta pH$ *or* $\Delta\,\tilde{\mu}_{H^+}$ and the flux represents metabolic turnover, such as consumption of oxygen, or production of ATP, or other metabolites. These relationships have been used systematically for the characterization of oxidative and photosynthetic phosphorylation, revealing some quantitative discrepancies with the tenets of the delocalised model, some of which can nonetheless be accommodated by assuming direct coupling between proton source and proton sink, in line with the proposals of the localised model. One outcome of such analysis has been an elaboration of the localised model termed the *mosaic protonic coupling* model, based on the hypothesis that the interaction between proton source and proton sink is characterized by their association into a number of independent functional units. (Westerhoff, 1984) While the identity of the functional units has been the subject of discussion, current understanding of mitochondrial physiology suggests that they may be associated to individual cristae or to smaller membrane domains.

### *1.22 Proton Mobility at the Membrane Surface*

The localised model is supported by several experiments that target the mobility of protons at the membrane-water interface of model systems, providing measures of interfacial proton conductivity and diffusion coefficients. The simplest model system, supported bilayers in air, already reveals the existence of measurable surface proton currents, of the order of 10-100 fA, as detected by Scanning Tunnelling Microscopy (STM) (Heim, 1995), which are modulated by atmospheric humidity. Surface proton flows can be observed in increasingly more complex mimics of biological membranes, such as at the lipid water interface of monolayers in a Langmuir trough (Teissié, 1985) and at submerged artificial phospholipid bilayers. Surface diffusion coefficients (D) for the proton are in the range $10^{-5}$ cm$^2$/s - $10^{-4}$ cm$^2$/s. (Serowy, 2003; Antonenko, 2008) The values are about one order of magnitude faster than the diffusion of lipid molecules, confirming that proton diffusion is not a simple manifestation of vehicle diffusion via associated lipid molecules. The same conclusions were obtained by using a flash of light to trigger proton release from the proton pump bacteriorhodopsin at the surface of its native membrane. (Heberle, 1994; Alexiev, 1995) The similarity of D values in the latter system, approximately $3x10^{-5}$ cm$^2$/s, with those obtained from artificial membranes suggests that the mechanism of surface proton diffusion is similar in all cases. Measurements of proton transfer rates towards the access channel of proton pumping proteins also support a role for the membrane in directing proton movement. (Ädelroth, 2004)

Theoretical investigations converge with experimental ones in identifying a preferential region for localised proton flow in the proximity of phospholipid headgroups. An adaptation of the Gouy-Chapman model of the electrical double layer shows that the extrusion of protons at a membrane leads to their accumulation in close proximity to the surface. (Koch 1986) Later studies confirmed that the region corresponds to a layer less than one nanometre in thickness, where protons are retained by an energy barrier attributed to the organization of interfacial water induced by membrane surface charges. The interfacial barrier limits diffusion of protons to the bulk phase and channels them along the membrane surface, where it is proposed that their mobility is mediated by proton hopping along an extended hydrogen bond network involving water and/or phospholipid headgroups. (Cherepanov, 2003; Serowy, 2003)

The following milestone, the measurement of proton flows in living cells, is a recent undertaking. Preliminary results have been reported using super-resolved microscopy techniques to image pH-sensitive subcellular probes in cellular mitochondria, although with conflicting conclusions. (Rieger, 2014; Toth, 2020) The area is nascent and more results are necessary to clarify outstanding issues and inconsistencies.

Overall experimental and theoretical reports over the past forty years have validated the basic assumptions of the localised proton model and have created conditions for its general acceptance.

### *1.23 Beyond Proton Flows. Alternative and Complementary Models*

While localised and delocalised proton conduction models have provided the currently accepted canon, alternative or complementary models of cellular energy management have been proposed by several investigators. Some of them can be considered as elaborations of the core concepts of the localised and delocalised models. Other proposals fully bypassed the requirement for proton flows, suggesting that transfer of energy takes place via direct interaction between membrane proteins. Among the latter, postulated mechanisms include specific interactions between source and sink mediated by conformational changes in the membrane (Boyer, 1977), by collision of the two (Slater, 1985), or by formation of long-lived complexes controlled by electrostatic interactions. (Tsong, 1987) Historically, these models received less attention than the ones based on proton gradients. Nonetheless, their interest is revived by the expanded role presently attributed to phase separation in biological membranes, whereby the associated compartmentalization can modulate protein-protein interactions. (Brown, 1998) Discussing them is beyond the scope of the present work, and the reader is directed to reviews that explored them in detail. (e.g. Nagle, 1986; Slater, 1987; Westerhoff, 1988)

## *2.0 Proton flows and proton circuits*

Solutions of electrolytes are conductors, whereby ions, including protons, act as charge carriers and allow the passage of current. The property was acknowledged by Mitchell in using the term *proticity*, as an analogue of electricity. (Mitchell, 1979) The connectivity of proton sources and proton sinks can be aptly represented in terms of circuit wiring. Proton currents can flow through an interface, such as a membrane, following proton wires in proteins, and they can flow along a membrane surface. In this picture, membrane associated structures can be modelled as electronic components. The hydrophobic core of the membrane acts as the dielectric of a capacitor, capable of storing energy in the form of charges on its opposite surfaces. Networks of hydrogen bonds are conductors, or semiconductors, that allow and direct proton flow. Proton pumps are equivalent to batteries, generating a voltage from chemical processes, or photodiodes, generating a voltage from the energy of absorbed photons. Proteins that use a proton current to produce work act as motors, including ATP-synthases and transporters.

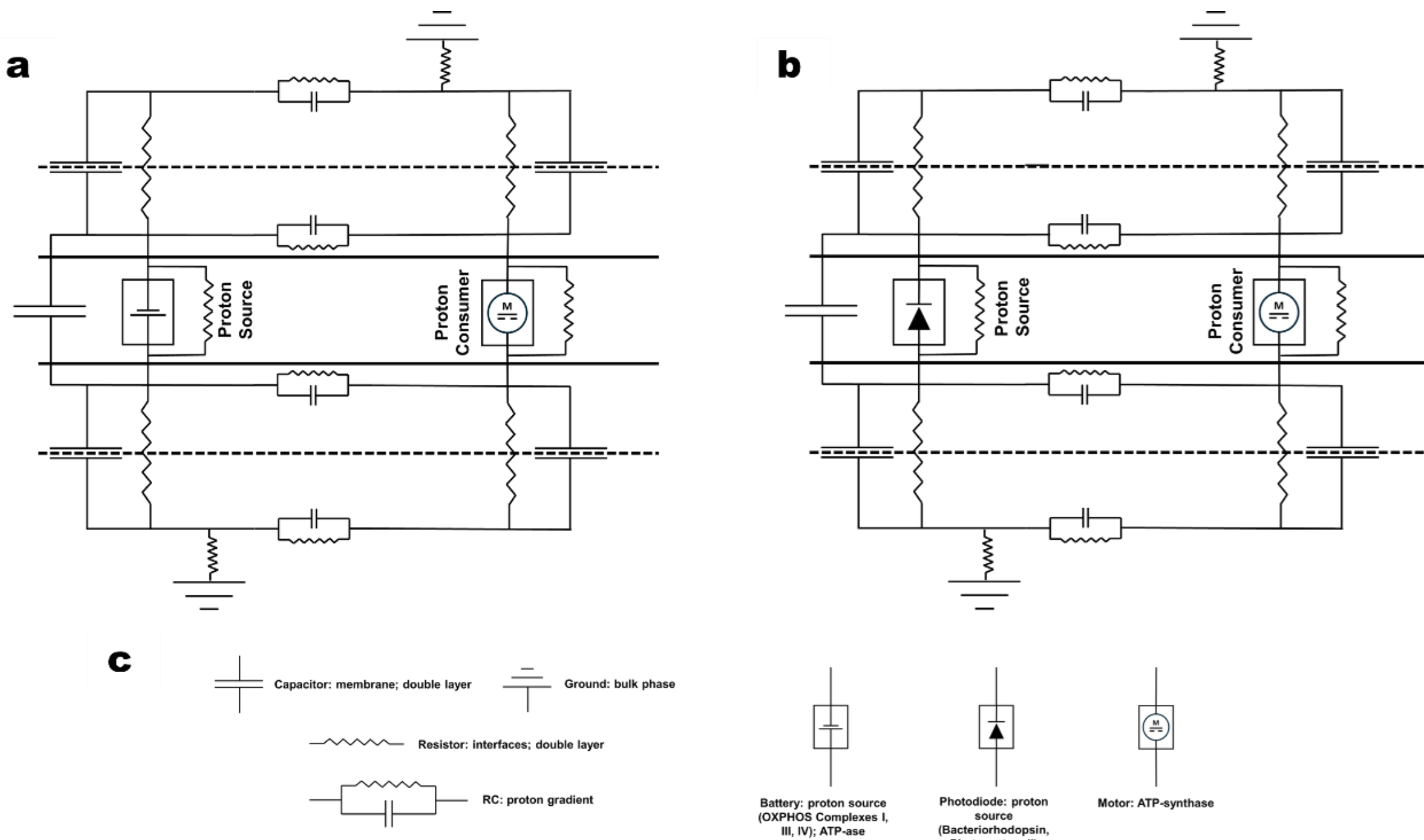

***Fig. 4 Electronic representation of proton circuits.*** *a: Oxidative phosphorylation. The proton source corresponds to a battery and comprises complexes I, III and IV of the mitochondrial inner membrane. The proton consumer is the mitochondrial ATP-synthase, equivalent to a motor. Reverse operation, as an ATP-hydrolase, would convert it to another battery. b:*

*Photosynthetic or halobacterial phosphorylation. The proton source is photosystem II, in the former case, or bacteriorhodopsin, in the latter; both of them are equivalent to a photodiode. The proton consumer is the archaeal or chloroplast ATP-synthase, equivalent to a motor. Also in this case, reverse operation, converts the protein to a battery. C: explanation of symbols*

Electronic circuit representations have been extensively used to model proton flows at a membrane. Figure 4 shows, as an example, a graphical representation of an elaboration of the localised model, describing multiple proton flows running parallel to a membrane surface. One channel for proton conduction is assumed in the proximity of the membrane, corresponding to the retention layer at the phospholipid headgroups. A second conduction channel is at the border of the electrical double layer region. Both channels comprise longitudinal proton gradients between source and sink. Because of the associated capacitance, they are equivalent to RC circuits. Transfer between the different channels is slowed by a resistance, while release of the protons to the bulk phase corresponds to transfer, and loss, of charges to the ground. The proton source is either a battery (Figure 4A), where voltage is produced by the terminal reactions in oxygen reduction, or a photodiode, energized by light absorption from bacteriorhodopsin or a photosystem (Figure 4B). The proton sink is an ATP-synthase, acting as a motor driven by the proton flow (Figure 4A). The latter can also be operated in reverse, as an ATP-hydrolase, thus acting as an additional battery that relies on hydrolysis of ATP to transfer positive charges. (Figure 4B).

Because the quantities that characterize the electronic components, such as capacitance and resistance, are determined by the geometry of the system, schemes such as the one in Figure 4 exemplify the close connection between the overall architecture of the membrane assembly and its performance in energy interconversion and storage. Historically, they have been valuable tools for testing and validation of hypothesis on the mechanisms of proton conduction.

### *3.0 Subcellular Architecture and Biological Energy Conversion*

Our current understanding of proton transfer at energized membranes has been greatly informed by the architecture of $F_1F_0$ ATP-synthase sites, where $F_1$ represents the globular subunit in the aqueous phase of the n-side and $F_0$ represents the transmembrane subunit. $F_1F_0$ ATP-synthases are present in all domains of life and provide the paradigm for the mechanism coupling proton flow to ATP synthesis. Detailed structural and functional information available in this respect comes from the eukaryotic enzymes located in mitochondria and chloroplasts, the organelles dedicated to energy production.

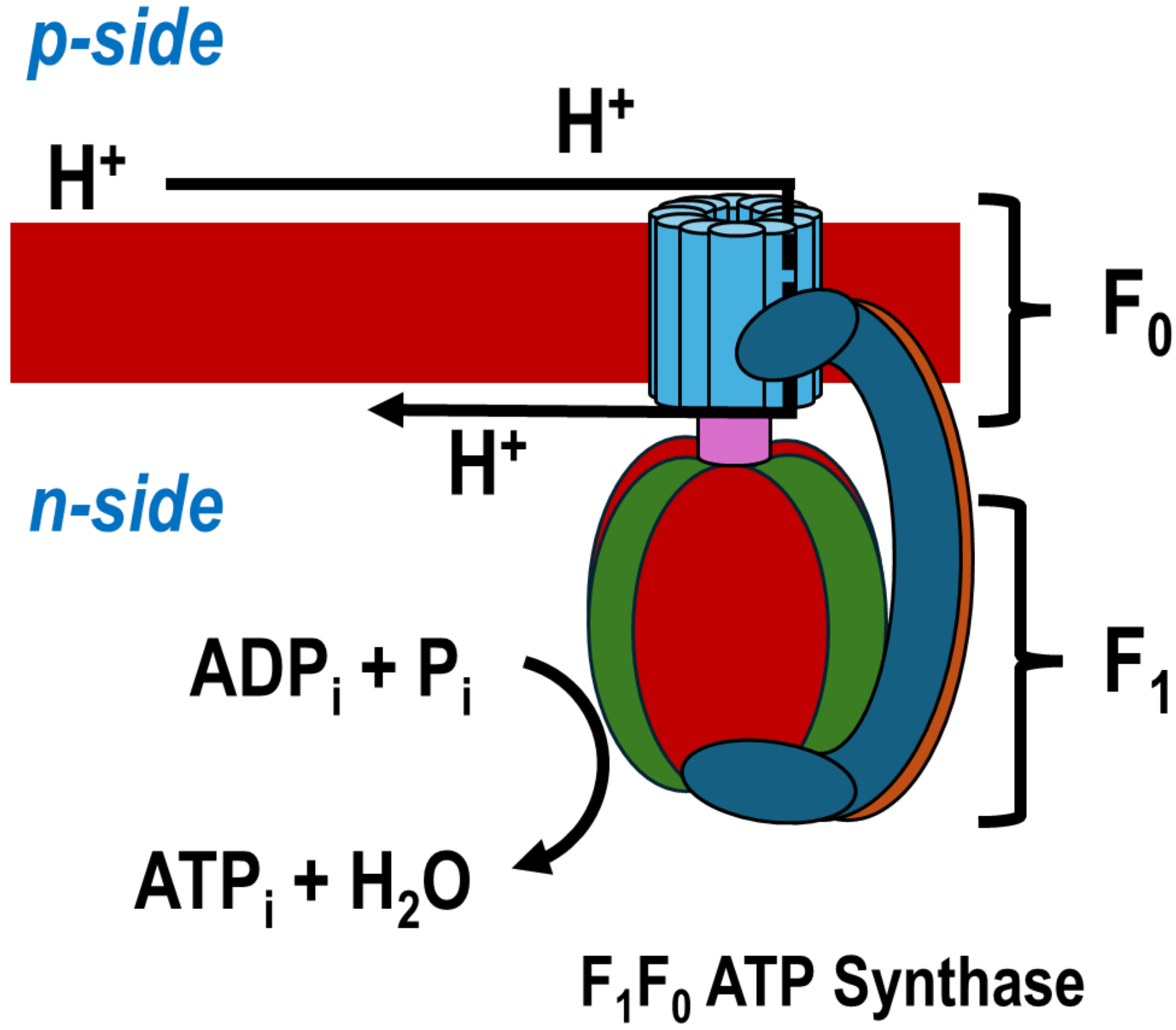

***Fig. 5 Structure of an $F_1F_0$ ATP-synthase***. *The soluble $F_1$ domain, exposed to the mitochondrial matrix, is the site of ATP-synthesis. Transfer of protons from the p-sid to the n-side across the $F_0$ domain rotates the axis of the protein and provides the energy for the conformational changes in $F_1$ that drive the synthesis of ATP*

ATP-synthases are considered the prototypical example of a molecular motor. They use the proton current along the potential difference within the membrane to generate ATP. (Kühlbrandt 2019) Figure 5 shows the schematics of an $F_1F_0$ ATP-synthase. ATP synthesis or hydrolysis occurs in the $F_1$ domain, within the aqueous phase of the mitochondrial matrix, while the proton current crosses the membrane by flowing through the hydrophobic $F_0$ domain. The two major domains are connected by ~~a~~ polypeptides acting as ~~a~~ the shaft or axis of the motor. An additional polypeptide spans the $F_0$ and $F_1$ domains, and has the function of the stator of the motor. (Walker, 1998) Proton flow across the membrane generates a torque that causes the rotation of the $F_0$ domain and the shaft relative to the $F_1$ domain, providing the energy for the conformational changes in $F_1$ that lead to ATP synthesis. (Boyer 1997) The transfer of protons across the membrane and along their concentration gradient releases the energy stored in the electrochemical potential difference, but does not directly supply protons for the condensation reaction in Equation 1, which are derived from the adjoining bulk phase. The number of protons transferred per ATP molecule synthesised is determined by the structure of the $F_0$ domain, instead of Equation 1, and is species dependent. (Kagawa, 2010; Nirody, 2020)

### *3.1 Mitochondrial ATP-synthases*

Mitochondrial ATP-synthases ($mF_1F_0$-ATP-synthases, or $mF_1F_0$-ATP-ases when working in reverse) are among the better characterized systems in terms of architecture of the protein-membrane assembly. The protein is located on the inner mitochondrial membrane, within the folded protrusions that extend into the matrix with the shape of flattened sacks, the cristae. The protein is oriented so that the p-side corresponds to the intracristae space, or lumen, with a thickness of approx. 20 nm. The lumen is connected to the space between the inner and outer mitochondrial membrane, the intermembrane space, via a narrow stem, the crista junction, which restricts the flow of matter between the two volumes. (Figure 6) ATP-synthases are arranged in ribbons of dimers located at the rims of the cristae. (Kühlbrandt 2019) It is proposed that the curvature of the membrane at the rim is imposed by the dimers themselves (Paumard, 2002; Strauss, 2008), which actively shape the geometry of the p-side. Regions of high membrane curvature, such as rims and apexes, favour the local accumulation of charges and it has been suggested that the geometry of the latter is tailored to channel proton transfer in the direction of the ATP-synthase. (Rieger, 2014; Davies, 2011) In contrast to the p-side, the n-side displays a more open environment, being turned towards the matrix side of the mitochondrion, with characteristic distances of 100 nm or greater (Table 1).

***Table 1.*** *Characteristic dimensions of the environment of eukaryotic ATP-synthases*

| ATP-synthase | p-side | n-side |
|---|---|---|
| **Mitochondrial** | ~20 x 500 x 500 nm$^3$ (disk intracristae space) | > 100 nm (mitochondrial matrix) |
| **Thylakoid** | ~4 x 300 x 300 nm$^3$ (thylakoid lumen) | > 100 nm (chloroplast stroma) |

### *3.2 Chloroplast ATP-synthases*

Chloroplast ATP-synthases ($cF_1F_0$ ATP-synthases, $cF_1F_0$ATP-ases) have the same overall architecture and mechanism as the mitochondrial $F_1F_0$ ATP-synthases, with a few differences in the number and identity of the subunits which account for differences in activity. They are located on the surfaces of thylakoids (Kühlbrandt, 2019) and in the chloroplasts of higher plants they appear to be mostly monomers that are randomly distributed on the portions of the thylakoid membrane exposed to the stroma. (Daum, 2010) The p-side of the membrane faces the thylakoid lumen, a fully enclosed compartment ~ 4.5 nm thick and with a lateral extension in the range 300-500 nm. The n-side faces the wider space of the stroma (Figure 6 and Table 1).

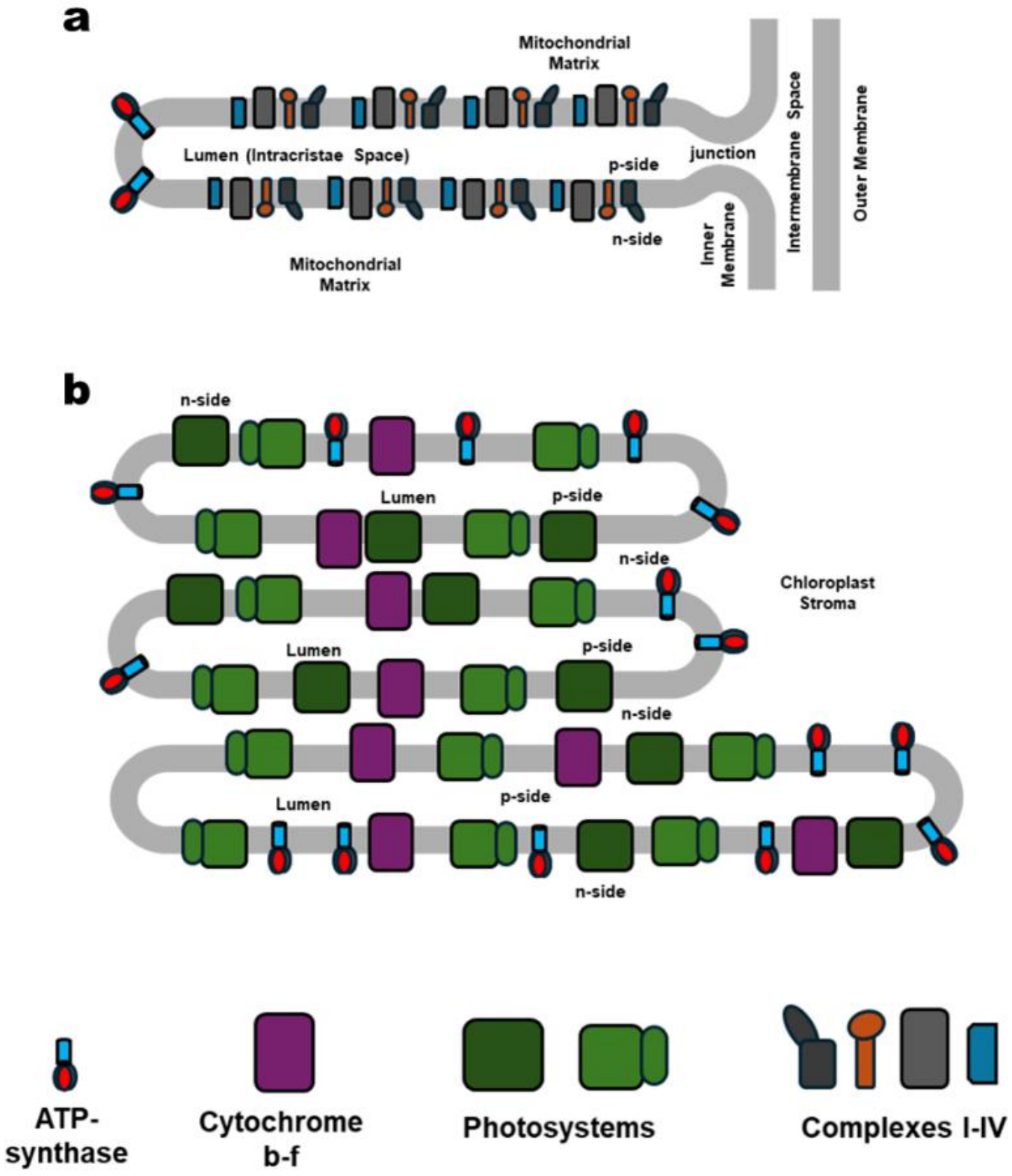

***Fig. 6 Distribution and topology of $F_1F_0$ ATP-synthase/ATP-ase enzymes in energy-conserving membranes**. a. Topology of the mitochondrial cristae membrane. The $mF_1F_0$ ATP-synthases are arranged in pairs and located at the rims of the crista. The complexes of the oxygen reduction chain are distributed in the flatter regions of the crista and pump to the intracristae space (p-side) protons from the mitochondrial matrix (n-side) where ATP synthesis or hydrolysis takes place. The crista junction connects the intracristae space to the space between the inner and outer mitochondrial membrane and controls flows between the two volumes. b. Topology of the chloroplast thylakoid membrane. The thylakoids are arranged in stacks inside the stroma of the chloroplasts. $cF_1F_0$ ATP-synthases are distributed on the outer side of thylakoid membranes that are directly exposed to the stroma, where ATP synthesis or hydrolysis takes place. The photosystems are distributed throughout the membrane and pump protons from the stroma and intra-thylakoid space to the thylakoid lumen*

### *3.3 Bacterial and Archaeal Systems*

Proton energetics in bacterial and archaeal organisms were the subject of intense early discussions and supplied some of the contentious issues that questioned the viability of the chemiosmotic model. (Krulwich, 1995) Bacterial and archaeal proteins were later employed extensively in the reconstituted systems used to assess localised proton flows. (Mulkidjanian, 2006) Despite the intense scrutiny, and the availability of detailed structural information for bacterial ATP-synthases and the proteins of the respective electron transport chains, the quantitative aspects of bacterial energetics still present abundant unresolved issues. The energy transducing membranes from these single cell organisms are not localised to organelles dedicated to energy production, but are constituents of the cellular envelope and support a wide range of processes, from signalling to metabolic turnover, that can affect proton concentrations *in vivo* and increase the complexity of the analysis. The area is open for future investigation, and discussing these aspects is beyond the scope of this review.

### *3.4 The Volume and Geometry of the p-side as Determinant Factors of Energization*

A common feature of $F_1F_0$ATP-synthase systems from energy producing organelles is the operation in environments that are geometrically well defined and locally constrained, particularly in the direction normal to the membrane on the p-side. The linear dimension normal to the membrane ranges from ~20 nm, in mitochondria, to about ~4.5 nm, in thylakoids. In the latter case it corresponds to a few water layers, and is just larger than the Debye length. The volume of the aqueous phase in contact with the p-side ranges from ~$10^{-4}$ $\mu m^3$ (thylakoid lumen) to ~$10^{-2}$ - $10^{-3}$ $\mu m^3$ (intracristae space), so small that the appropriateness of describing associated processes with macroscopic quantities has been questioned in early discussions. (Mitchell, 1967a; Mitchell, 1967b) The limitations of macroscopic theoretical treatments in accounting for the small dimensionality of these structures could be responsible for some of the inconsistencies reported for the chemiosmotic theory, which relies on bulk quantities to define Equation 2.

The volume of the p-side compartment affects the contribution of ΔpH to the total *pmf.* Its dimensions imply that the addition of a relatively small number of protons corresponds to major changes in nominal concentration and in thermodynamic properties. Increasing the amount of free $H^+$ by a single ion in the thylakoid lumen, with a volume of ~$10^{-4}$ $\mu m^3$, corresponds to a ~$10^{-6}$ M increase in concentration, which can lower the pH of the compartment from pH = 7 to pH ~ 6. The same amount of free protons in the mitochondrial intracristae space would produce a drop in pH from pH = 7 to pH ~ 6.8. Ultimately, the equilibrium concentration of free $H^+$ ions in the p-side compartment is determined by its buffering capacity, which includes contributions from aqueous phase buffers and membrane molecules, as recognized since the early descriptions of chemiosmosis. (Uribe, 1968; Mitchell, 1969) Together with the permeability of the membrane to protons and other cations, these factors can account for the different contribution of $\Delta pH$ to $pmf$ in the two compartments.

One notable consequence of this compartmentalized geometry is that the criticism originally raised towards the chemiosmotic model, about the unrealistically long time needed for a proton to diffuse from membrane to bulk and the resulting negligible change on $\Delta\,\mu_{H^+}$ (the Pacific Ocean analogy) is not applicable to contained nanoscale volumes, such as the mitochondrial and thylakoid systems. Assuming D ~ $10^3$ $\mu m^2/s$, the same as for in-plane diffusion, and neglecting the presence of barriers for surface to bulk diffusion, a proton can cover the transversal length of the lumen in 0.1 μs to 1 μs. Diffusion times decrease by more than one order of magnitude if the value of D for proton diffusion in bulk water is used (D ~ 9 x $10^4$ $\mu m^2/s$), fast compared to the characteristic time for proton transfer to the ATP-synthase, of the order of $10^{-3}$ – $10^{-4}$ s.

Another consequence of the small dimensions and geometry of eukaryotic p-side compartments is that their properties are consistent with aspects of both the localised and delocalised proton models. Protons can rapidly diffuse through and equilibrate in the aqueous phase of such compartments, and store energy via changes in electrochemical potential, in agreement with the description of the delocalised proton model. At the same time, the highly organized molecular scale environment allows for direct proton transfer between source and sink, modulating the energy transfer process, in agreement with the localised model description. The difference between the two models lies in their description of different facets of the same world. Using the language of protic circuit representations, the delocalised proton model describes energy storage at equilibrium by the battery of the circuit, while the localised model describes charge flow along the wiring of the circuit.

## *4.0 Conclusions and Future Perspectives*

Current structural information on the environment of ATP-synthases *in vivo* supports specific aspects of both dominant models of cellular energy management, based on localised and delocalised protons. In contrast to the early competitive relationship, the models increasingly appear to be complementary descriptions of separate facets of cellular energy management. While the delocalised model focuses on energy balance at thermodynamic equilibrium, the localised model and its derivatives address the mechanistic aspects of energy conversion out-of-equilibrium. Other models of the mechanism of energy transduction have also been proposed over the years, then abandoned in favour of the latter two. In light of the large amount of structural information now available on the compartments that regulate ATP synthesis and hydrolysis, it is now worth reconsidering if any aspects of such alternative models can be recovered and incorporated in a more exhaustive description of energy transduction. The arbiter of such synthesis will be the quantitative validation of hypothesis in single living cells and functional organelles. Performing such measurements involves reporting concentration dynamics in intact structures with subcellular and sub-organellar resolution and will be the challenge for the foreseeable future. To date such measurements have been performed by imaging GFP markers with super resolved fluorescence techniques (Rieger 2014; Toth 2020; Wolf 2019), which involve molecular engineering of the cellular system. Minimizing or avoiding perturbation of the system would be highly desirable, and must complement the need for high spatial resolution, sensitivity and time resolution. It is left to the ingenuity of experimentalists to overcome these limitations by combining existing tools into appropriate experimental design.

[1] This is a reprint of Mitchell's review of chemiosmotic coupling, known as the first "Grey Book", originally published in 1966 as Research Report 66/1 by Glynn Research Ltd.